\documentclass[useAMS,usenatbib]{mn2e}
\usepackage{aas_macros}
\usepackage{graphicx}
\usepackage[usenames]{color}
\usepackage{times}
\def\gtsim{\lower.5ex\hbox{$\; \buildrel > \over \sim \;$}}

\title[CMB Cold Spot Redshift Survey]{A redshift survey towards the CMB Cold Spot}

\author[ M.N. Bremer et~al.]{M.N. Bremer$^1$, J. Silk $^2$, L.J.M. Davies$^1$, \&  M.D. Lehnert$^3$ \\
$^1$H H Wills Physics Laboratory, Tyndall Avenue, Bristol, BS8 1TL, UK\\
$^2$
Department of Physics,
Denys Wilkinson Building,
Keble Road,
Oxford, OX1 3RH, UK\\
$^3$Laboratoire d'Etudes des Galaxies, Etoiles, Physique et Instrumentation GEPI, UMR8111,  Observatoire de Paris, Meudon, 92195 France\\
}

\begin{document}

\date{Accepted . Received ; in original form }

\pagerange{\pageref{firstpage}--\pageref{lastpage}} \pubyear{}

\maketitle

\label{firstpage}

\begin{abstract}
We have carried out a redshift survey using the VIMOS spectrograph on
the VLT towards the Cosmic Microwave Background cold spot. A possible
cause of the cold spot is the Integrated Sachs-Wolfe effect imprinted
by an extremely large void (hundreds of Mpc in linear dimension) at
intermediate or low redshifts. The redshift distribution of over seven
hundred $z<1$ emission-line galaxies drawn from an $I-$band flux
limited sample of galaxies in the direction of the cold spot shows no
evidence of a gap on scales of $\Delta z\gtsim 0.05$ as would be
expected if such a void existed at $0.35<z<1$. There are troughs in
the redshift distribution on smaller scales ($\Delta z\approx 0.01$)
indicating that smaller scale voids may connect regions separated by
several degrees towards the cold spot. A comparison of this
distribution with that generated from similarly-sized subsamples drawn
from widely-spaced pointings of the VVDS survey does not indicate that
the redshift distribution towards the cold spot is anomalous or that
these small gaps can be uniquely attributed to real voids.

\end{abstract}


\section{Introduction}
\label{sec:intro}

A major triumph of modern physics is the measurement and explanation
of the Cosmic Microwave Background (CMB) fluctuation spectrum ({\it
  e.g.} Bennett et~al., 2003).  Encoded within the
pattern of fluctuations from constant temperature on the sky are the
signatures of the makeup of the Universe. The amount of baryons,
matter, dark energy within the Universe, when and how reionization
occurred, and the power spectrum of initial fluctuations can all be
inferred either from the properties of these fluctuations alone or in
combination with other measurements.

\begin{figure}
\includegraphics[width=8.5cm]{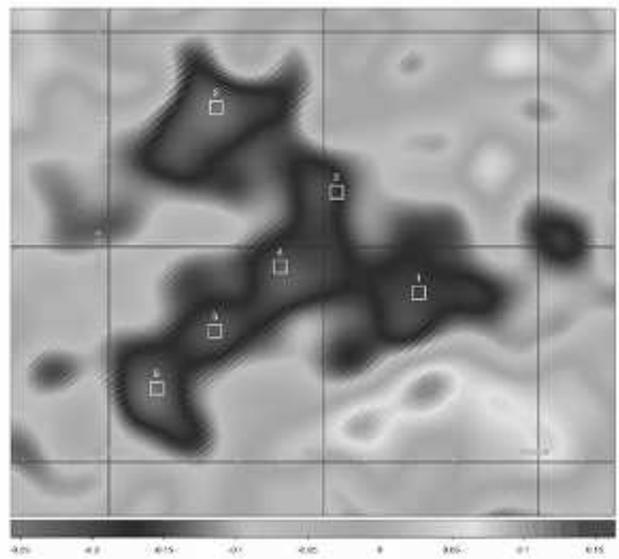}
\caption{The six spectroscopic field centres overlayed on the WMAP ILC temperature map of the cold spot region. The pointing centres are: (1) 03:28:51 -21:03:54,
(2) 03:10:00 -16:45:44, (3) 03:21:13 -18:43:15, (4) 03:15:57 -20:27:23, (5) 03:09:53 -21:57:24
and (6) 03:04:29 -23:17:10}
\label{fig:1}
\end{figure}

The ``concordance cosmology'', based strongly on $\Lambda$CDM is
extremely successful in explaining CMB anisotropies on small angular
scales. However, there are clear anomalies on larger angular scales
which pose potential problems for our understanding of the evolution
of structure at the most fundamental levels. The most striking anomaly
is the presence of an apparent cold spot in the Wilkinson Microwave
Anisotropy Probe (WMAP) CMB data in the southern hemisphere (Vielva et
al., 2004; Cruz et al., 2005; Spergel et
al., 2007). This region of decreased temperature
($\Delta T/T \sim -10^{-5}$) extends over an angular scale of
5-10$^\circ$ (Figure 1). The cold spot has a less than 2 per cent
probability of being generated as part of a Gaussian fluctuation
spectrum (Cruz et al., 2007a,).

\begin{figure*}

\includegraphics[width=18cm]{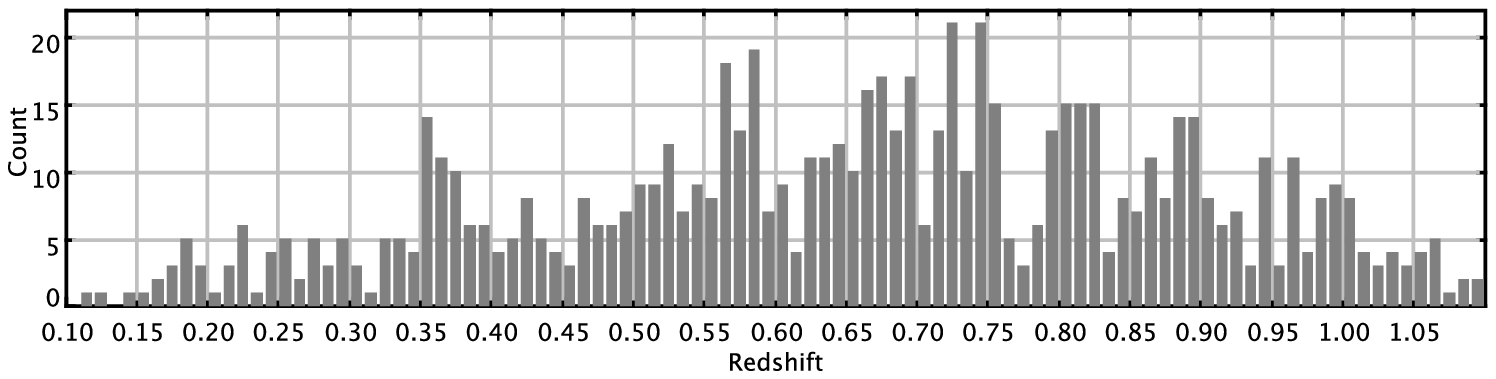}

\vspace{-1cm}

\includegraphics[width=18cm]{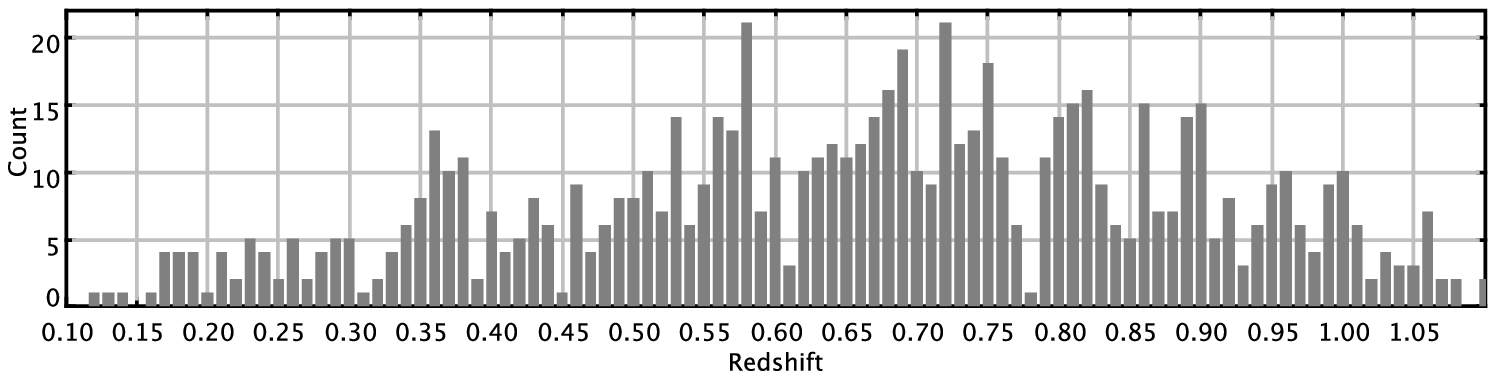}
\vspace{-1cm}
\caption{The redshift distribution of the sources in the six
  fields. Each bin has a width of 0.01 in redshift and the
  distribution is displayed twice, the second time with the the binning
  offset by half a bin width. The expected signature of a
  self-compensating void large enough to produce the cold spot
  temperature decrement through the ISW effect is an absence of
  redshifts across several consecutive bins and sharp spikes in the
  distribution at either edge of the gap. No such feature is seen.}
\label{fig:2}
\end{figure*}

Assuming that the cold spot is not a statistical artifact, or produced
as a result of the way the data has been analysed ({\it e.g.}  Zhang
\& Huterer, 2009), determining the nature of the cold spot is crucial
to our understanding of the growth of structure. Either it is
intrinsic to the CMB itself or is imposed by a foreground structure along
the line of sight to the CMB. Either way, there are important
implications for our understanding of structure evolution. In the
first instance, an intrinsic origin points to extra physics (such as
textures or defects, Cruz et al., 2007b, 2008; Durrer 1999) or non-standard
inflation ({\it e.g.} Peebles 1997) influencing the behavior of the
early Universe.

A Galactic origin for a foreground source of the cold spot appears to
be ruled out by its spectrum (Cruz et al., 2006). If
the cold spot is a secondary CMB anisotropy caused by modulation of
the CMB signal as it passes through an intervening massive structure,
there is only one possible cause.  An extremely large void of scale
$\sim 200 h^{-1}$Mpc could cause the signal through the Integrated
Sachs Wolf (ISW) effect (Inoue \& Silk 2006).  As the
CMB photons pass through such a void, over their transit time the
gravitational potential of the void evolves as the void expands
leading to a net redshift of the photons and hence cooling of the
spectrum at that point in the sky. The size of the void is set not
just by the angular size of the cold spot, but also by the light
travel time across the void necessary to induce an ISW signal of this
magnitude. Any void must be at $z<1$ as at later times the increasing
influence of the cosmological constant on the expansion of the
Universe enhances the ISW signature of the void.

We have carried out a redshift survey of galaxies selected from six
patches of sky separated by $\sim 1-2^\circ$ towards the CMB cold spot
in order to search for the signature of an extremely large void
capable of producing a large enough ISW signal to account for the CMB
feature. Assuming galaxies in such a structure have bias factors
similar to those found in typical voids, this structure should be largely
empty of galaxies, particularly those around $M^*$ which dominate the
galactic mass distribution. Given its extraordinary size (2-300
Mpc, of order 5-10 percent the distance back to $z\sim 1$), it should
leave a clear gap of $\Delta z \sim$ few $\times 0.01$ in the redshift
distribution towards the cold spot depending upon the exact size and
redshift of the void.

\section{Observations and data reduction}
\label{sec:observations}

Six separate pointings towards the cold spot were imaged in the
$I-$band using VIMOS (LeFevre et~al. 2003) on the VLT in service mode
between August and October 2008 (ESO service run 082A-0367(A)).  The
six pointings, shown in figure 1 were chosen to sample areas within
the region defined as the cold spot in the WMAP ILC data while
avoiding optically bright objects. The data were obtained in
photometric conditions with seeing of typically 0.8 arc-seconds. Five
dithered exposures totaling 1400 seconds were obtained in each
pointing. The data pipeline bias subtracted and flat-fielded the images. Each
image was then de-fringed. The five images for each detector quadrant
of each pointing were aligned and combined into a final image. A
catalogue of all of objects in each image was obtained using
Sextractor (Bertin \& Arnouts 1996) and the zero-point appropriate for
each night was applied. The number counts in the catalogues typically
turned over between $24.3<I<25$.

\begin{figure}
\includegraphics[width=8.5cm]{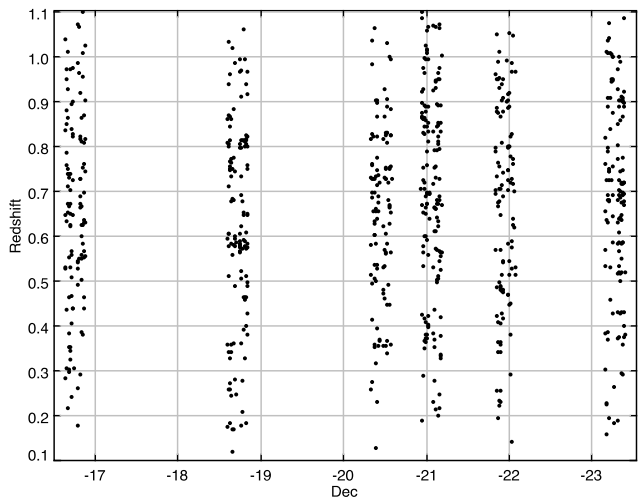}
\caption{ A plot of the Declination and redshift of each source contributing to
  figure 2. There is no coherent large gap in the redshift
  distribution common to a subset of the fields that could be
  attributed to a void large enough to produce the cold spot
  signature. }
\label{fig:3}
\end{figure} 

These catalogues were then used to design the spectroscopic masks. As
the coordinates used for VIMOS mask design are defined through the
$R-$band filter, the coordinates in the catalogues were transformed
using a (first order) shift and stretch in each of X and Y, determined
by a comparison of positions of common objects in the $I-$band images
and a reference $R-$band image. Stretches were typically less than a
pixel (0.2 arcsec) across an entire quadrant.  A subset of objects
with $21.5<I<22.8$ were selected from these catalogue as input into
the VMMPS mask-making software (Bottini et~al. 2005). Two sets of
masks were made for each pointing with between 41 and 70 slits on each
quadrant. In total, 2768 objects were essentially randomly targeted
for spectroscopy from a total sample of just over 8600 objects in the
magnitude range.

Spectroscopic observations were carried out through these masks from
November 2008 to February 2009 and during October 2009 (ESO service
run 082A-0367(B)). Each mask was observed for 9900 seconds split into
9 exposures with the objects moved along the slits between them.
Slits were 1 arcsecond wide and at least 8 arcseconds long. The LR-red
grating along with the OS-red order sorting filter was used to give a
spectral coverage between 5500 $\AA$ and 9500$\AA$. Strong fringing
beyond $\sim 7600\AA$~ limited the utility of the red part of the
spectra.  The resolution of the final spectrum was typically $R=210$
and this was sampled by at least three spectral pixels.

The data were reduced using a bespoke IDL-based pipeline. Each frame
was bias subtracted and flat-fielded. Each spectrum within a frame was
then sky-subtracted and resampled onto a common wavelength scale. The
nine separate exposures for a given mask were then combined and the
sky subtraction re-performed to remove any remaining residual sky
emission.  The multiplexed design of the spectrograph is such that the
zeroth-order spectra of some slits fall onto the main first-order
spectra of another. This results in either a region of the spectrum
with extremely high noise, or an apparent broad emission line in
either the sky or on the main object if the compressed zeroth-order
spectrum of the other object falls onto that of the main target. These
regions were automatically flagged to avoid misleading results. 

The spectrum of each object was then optimally extracted from each
two-dimensional spectrum. Its wavelength calibration was
double-checked by cross-correlating the slits wavelength-calibrated
sky spectrum (which was propagated through the pipeline) to a
reference sky spectrum. Cases where there were differences of more
than 2$\AA$ were noted. The spectrum was multiplied by the standard
spectral response for the instrument configuration which gave a
sufficiently good spectrophotometric calibration for the purposes of
redshift determination

Each spectrum was examined by eye. For the purposes of this project
redshifts were only assigned for objects with emission lines. Each
object was given one of four classifications: (1) secure redshift, (2)
very probable redshift (3) possible redshift and (4) no redshift. In
the first case multiple emission lines or a clear line plus other
spectral features needed to be present. Typically this would be the
[OIII] doublet in combination with the H$\beta$ line, or an [OII] line
in combination with [NeIII] or CaH\&K. Very rarely the H$\alpha$/NII
lines were identified along with the [SII] doublet. In the second case
either a strong single line was detected or a second potentially
confirming feature was present but compromised by fringing or large
sky variation in the red.  Based on published equivalent width
distributions and the relative volumes probed at $z<0.2$ and $z>0.75$
(where $H\alpha$ and [OII] could be confused given the resolution of
the spectra), single clear lines were attributed to [OII] rather than
H$\alpha$. In the third case a line was detected, but attribution was
considered insecure as the strength of the line was insufficient ({\it
  e.g.} the line could plausibly be the strongest of the [OIII]
doublet lines while leaving the weakest undetected). For the following
we only consider the highest two classifications.  The spectral
resolution was such that centroided wavelengths of individual lines
from these objects could be determined to an accuracy of better than
10 Angstroms (so redshifts to better then 0.0015 in the absence of other
systematic effects ).

\begin{figure*}
\includegraphics[width=18cm]{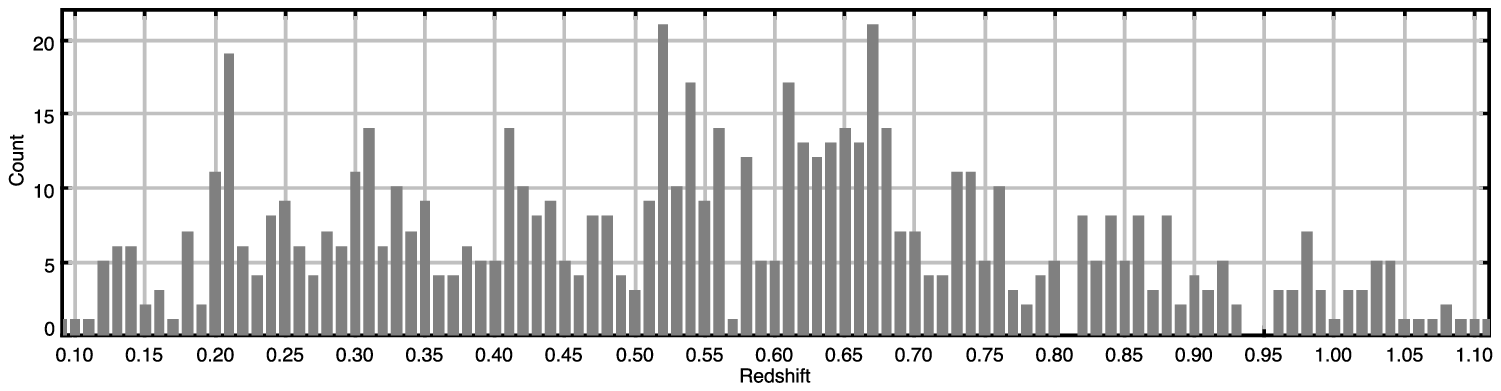}
\includegraphics[width=18cm]{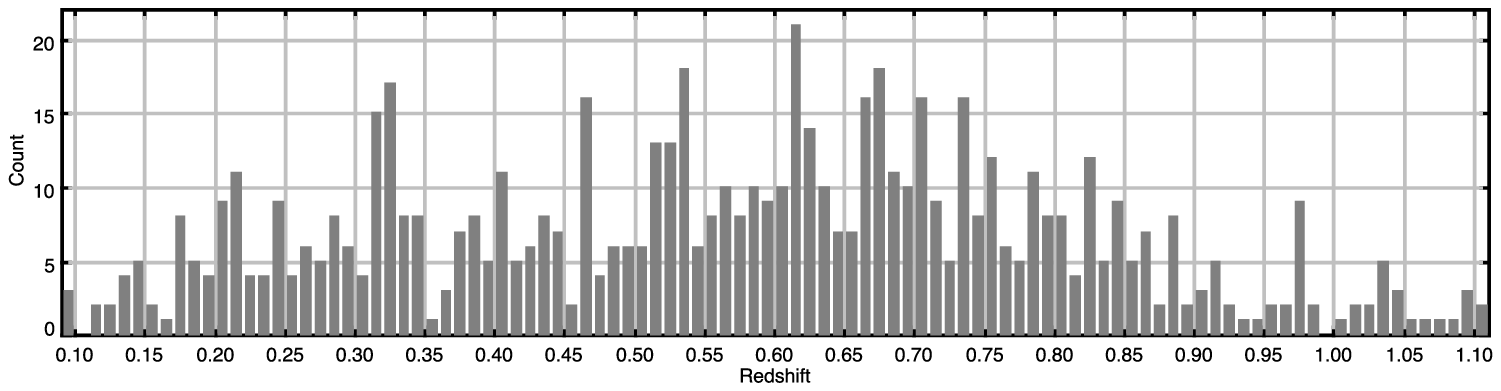}
\caption{Two example redshift distributions of samples of VVDS sources
  selected in a similar way and in a similar sample size to our survey
  and generated from four widely-spaced pointings on the sky. These
  distributions show redshift bins empty or almost empty of sources,
  indicating that similar gaps in the redshift distribution of our
  survey cannot be securely attributed to coherent structures across
  all or the majority of our pointings. In the case of these VVDS
  redshift distributions, the distance between the four pointings are
  so large that no coherent structure can connect them.}
\label{fig:4}
\end{figure*}

In total 750 objects (just under 10 per cent of the total flux limited
sample) were classified as having secure or very probable redshifts,
734 of these at $z<1.1$. The highest redshift object was a quasar at
$z=4.614$. Most objects at $z<0.5$ were primarily identified through
their [OIII] emission, and those at $0.5<z<1.1$ primarily from [OII]
emission. Although these could in principle be identified from [OIII]
emission out to $z\sim 1$, the strong fringing and relatively poor red
sensitivity of the detectors meant that the [OIII] doublet was not
the primary means of redshift confirmation beyond $z\sim 0.5$.

\section{Redshift Distribution}

The redshift distribution of the 734 objects at $z<1.1$ is presented
in figure 2. The distribution is shown for a bin size of $\Delta z
=0.01$ and duplicated having shifted the bin centres by half a bin
width. The signature of a $\sim 300$Mpc self-compensating void is a
gap (or at least a significant lack of redshifts) in a region of size
$\Delta z \sim 0.05$ or more, and spikes in the redshift distribution
either side of that void, from the ``walls'' of galaxies that mark out
the perimeter of the structure. No such features are seen in this
redshift distribution. At $z<0.35$ the comparatively small number of
objects in this distribution may mean that such a structure could be
hidden by small number statistics, but out to $z<1$ where there are
several tens or over a hundred of galaxies per 0.1 in redshift, the
lack of an obvious gap or strong deficit in the distribution is clear
and significant. This data set rules out an unusually large void
connecting these six pointings. The redshift distributions of the
individual pointings were examined in order to determine whether a
large void could be present in a subset, see figure 3. Again, no
obvious gap of significant size exists in common across a subset of
the pointings.

\begin{figure}
\includegraphics[width=9cm]{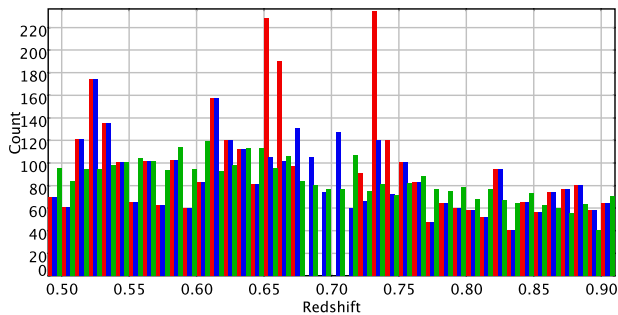}

\caption{ Redshift distribution of reliably-confirmed sources in VVDS
  (blue). Imposed on this is the faked signature of a huge
  self-compensating void at $z=0.7$(red). In green is the same
  redshift distribution, but with each redshift randomised by the
  addition of a number drawn from a Gaussian distribution with
  $\sigma=0.05$. The signature of the void is removed by the effective
  smoothing caused by this randomisation, which is of the same size as
  typical photometric redshift uncertainties.}
\label{fig:5}
\end{figure}

On smaller scales there are several redshift bins that show a deficit
of sources across all pointings in one or both of the binnings. Only
one pointing contains an object with a redshift in the range
$0.775<z<0.785$ and so this could potentially represent a coherent void
across a large area of sky, albeit one that was a sheet-like system
oriented in the plane of the sky. However, it could also be that this
deficit is purely down to the statistics and there is no coherent
structure connecting the pointings.  To test for this we can generate
a comparison sample from spectroscopy of VVDS objects. This
spectroscopy was obtained using VIMOS in essentially the same
configuration as was used for our survey. 

The publicly available VVDS data includes spectroscopic catalogues for
three completely independent pointings on the sky separated by tens of
degrees (VVDS-CDFS, VVDS-DEEP and VVDS-F22), with spectroscopy in all
three regions reaching to at least $I_{AB}=22.5$ (Lefevre
et~al. 2004,2005) . We selected four areas the size of our individual
points, one each from VVDS-CDFS and VVDS-DEEP and two (separated by
over a degree) from the larger area VVDS-F22 field. We selected 175
galaxies from each of these four areas, drawn from sources flagged as
either $>90$ per cent secure redshifts or single line redshifts, the
closest match to our selection. We repeated this process several
times in order to generate sample redshift surveys with approximately
the same number and quality of redshifts as in our survey, albeit
drawn from four as oppose to six patches of sky.  Three of the four
patches are so far apart no coherent 3-dimensional structure could
connect them. Of the two that were closer together, there was no
expectation of such a structure connecting them, the posited void
towards the cold spot ought to be a unique structure on that scale.

Examples of  redshift distributions for these samples are plotted in Figure
4. As the redshifts for the VVDS survey are generated initially by an
automatic pipeline, the overall shape of the distribution is slightly
different to that of ours. This is not significant for our
purposes. What is clear is that the redshift distribution of samples
drawn from essentially random areas of sky also show empty or near
empty redshift bins on scales of $\Delta z=0.01$. As these are
widely-separated fields, these cannot represent true
spatially-coherent structures that connect the individual fields, but
must simply reflect the statistical variation in the redshift
distribution for multiple-field surveys of this size. Given that the
kinds of objects targeted in these samples are comparable to those in
our survey, those gaps that are observed in our cold spot redshift
distribution cannot be securely attributed to real voids that extend
over several of the pointings in our survey.

\section{Discussion}

The area of sky towards the cold spot has been examined
photometrically in several wave-bands other than in the mm-wave in oder
to search for the signature of a super-void. Rudnick et~al (2007)
claimed that there is a 25-40 per cent deficit of NRAO VLA Sky Survey
(NVSS) sources towards the cold spot and McEwen et~al. (2007)
demonstrated that the spatial correlation between the NVSS number
counts and fluctuations in the WMAP data are strongly influenced by
the cold spot area. This is puzzling given the expected redshift
distribution of NVSS sources ({\it e.g.} Raccanelli et~al. 2008,
Brookes et~al. 2008), which is broad and peaks at $z \sim 1$. Any void
at $z<1$ and with a size of $\Delta z \sim 0.1$, even if empty of
radio sources and non-compensating for these objects, should depress
the number counts towards the void by only a few per cent given the
volume of the void and the effective volume probed by NVSS (see also
Smith \& Huterer 2008).

Conversely, a deep optical imaging survey carried out by Granett,
Szapudi \& Neyrinck (2009) showed no evidence for a super-void towards
the cold spot. Using photometric redshifts derived from $g,r,i,z$
imaging of seven fields within the cold spot region (their pointings
A, C and D cover our pointings 6, 2 and 4, and their F \& G partially
contain our 3 and 1), they claimed to rule out any 200 Mpc void
between $0.5<z<0.9$. While this agrees with our spectroscopic
measurements, we note that attempting to identify a self-compensating
void with photometric redshifts which have uncertainties larger than
the expected void signal is challenging. This assumes that the
uncertainties in photometric redshifts can be fully captured by
Gaussian-like error kernels. In reality, while photometric redshifts
can be calibrated against the spectroscopic redshifts of objects
amenable to spectroscopy, it is not clear that those objects
unsuccessfully targeted by spectroscopy (often, as in our case the
majority of the targeted objects) follow this calibration. Thus
photometric redshift uncertainties could be considerably larger than
those characterised from a simple comparison to the successful
spectroscopy.

To illustrate the difficulty of using photometric redshifts to probe a
putative large self-compensating void, we took the spectroscopic
redshifts of all $I_{AB}<22.5$ VVDS sources with reliable or
single-line redshifts (over 7300 objects at $z<1.4$), changed the
redshifts of sources at $0.675<z<0.725$ to be at the limits of this
range, thereby simulating a self-compensating void with $\Delta z\sim
0.05$. The redshifts were then randomized by shifting each by a random
number drawn from a Gaussian distribution of width $\sigma=0.05$,
simulating the statistical uncertainties of photometric redshifts (but
ignoring the systematic effect noted above). Figure 5 shows the three
redshift distributions: the original VVDS distribution, the same but
with the imposed fake void, and then this distribution with the
simulated photometric uncertainties. As can be seen, even a broad,
sharp void, one with the strongest possible signature for any cold
spot, produces no discernible signal in the ``photometric'' redshift
distribution because of the smoothing effect of the uncertainties.

Given the ambiguities of the photometric treatments, the advantage of
the spectroscopic approach used here is clear. Our results show no
evidence of a void connecting the six pointings towards the cold spot
that is large enough to generate the cold spot through the ISW
effect. Given the overall redshift distribution of the galaxies with
confirmed redshifts, it is possible that such a void could exist at
$z<0.35$ and with typically only 1-4 sources per $\Delta z=0.01$ bin
in the absence of a void, we would not be sensitive to it. Any future
spectroscopic analysis of this area of sky should concentrate on the
region at $z<0.5$ The amplitude of the secondary CMB fluctuation cold
spot in the CMB is $7\pm 3.10^{-5}$ (Cruz et al. 2007b), and if due to
the Rees-Sciama effect, varies as the cube of the void radius. Hence
the uncertainty in cold spot amplitude at specified angular extent
translates into roughly a 15 per cent uncertainty in redshift. Consequently
while any void is unlikely to be at $z < 0.5$, this possibility cannot
be excluded.

Could we have missed a large void at $0.5<z<1$ through the placement
of our fields?  Our fields were chosen based on the WMAP ILC III map
of the cold spot region and cover several percent of the region with a
decrement larger than 100 $\mu$K.  If any void had a covering (and
filling) factor of less than 100 per cent in this region and our field
placings were essentially random within this, then there is a
possibility that we could have missed the void. The covering factor
cannot be much below 50 per cent as the ISW decrement is directly
proportional to the covering factor. For a given signal, any decrease
in the covering factor would have to be compensated for by an
increased depth to a void in the line-of-sight, leading to potentially
implausible geometries for the structure. If an under-density towards
the cold spot was made up of several smaller voids with a total
covering factor of 50 per cent rather than a single structure, the
combined ISW effect from these will produce a far smaller (and
therefore undetectable) temperature decrement than a single large
void, and so such a scenario is very unlikely. Given the arrangement
of our fields across the cold spot region, it is difficult to see how
all or even a majority could have missed a single large system.

Our results indicate a large void at $0.5<z<1$ is unlikely to have
generated the CMB cold spot temperature decrement through the ISW
effect. Either such a void is at considerably lower redshift than
expected, or there is some other explanation for the cold spot, one
possibly involving new physics or a different analysis of current and
future CMB data.

\section {Acknowledgments}

This work is based on Based on observations made with ESO Telescopes
at the La Silla and Paranal Observatory under programme ID
082-0367. Extensive use was made of the STILTS and TOPCAT software
packages (Taylor 2005).

\label{lastpage}

\end{document}